\documentclass[aps, prb, reprint, showpacs]{revtex4-1} 

\usepackage{amsmath}
\usepackage{graphicx}
\usepackage{amsmath}
\usepackage{amssymb}
\usepackage{url}
\usepackage{bm}

\begin{document}

\title{Long-distance radiative excitation transfer between quantum dots in disordered photonic crystal waveguides}

\author{Momchil Minkov and Vincenzo Savona}

\affiliation{Institute of Theoretical Physics, Ecole Polytechnique F\'{e}d\'{e}rale de Lausanne EPFL, CH-1015 Lausanne, Switzerland}

\date{\today}

\begin{abstract}
We theoretically investigate the magnitude and range of the photon-mediated interaction between two quantum dots embedded in a photonic crystal waveguide, including fabrication disorder both in the crystal and in the dot positioning. We find that disorder-induced light localization has a drastic effect on the excitation transfer rate -- as compared to an ideal structure -- and that this rate varies widely among different disorder configurations. Nevertheless, we also find that significant rates of $50 \mathrm{\mu eV}$ at a range of $10 \mathrm{\mu m}$ can be achieved in realistic systems.  
\end{abstract}

\pacs{78.67.Hc, 42.50.Ct, 42.70.Qs, 03.67.-a}

\maketitle

Semiconductor quantum dots (QDs) have very recently become candidate building blocks of a quantum information technology, after the experimental proof of full single-qubit control\cite{Patton_2005, Press2008, Berezovsky_2008, Greilich_2009, Greilich_2011, Poem_2011, Muller_2012, Godden_2012}. Beyond that, the possibility for \textit{two} qubits to interact coherently in a controlled fashion is an essential requirement for two-qubit quantum gates, that are a building block of the mainstream quantum information protocol.\cite{Nielsen2004} Given the localized nature of the quantum dots, a quantum bus is needed to provide the link between distant QD qubits\cite{Kimble2008}. In a semiconductor system, photons are an obvious choice for this task, due to their weak coupling to the environment (long decoherence time), and long-distance propagation. Additionally, semiconductor Photonic Crystal (PHC) devices have advanced to a remarkable level of sophistication. The state-of-the-art sub-nanometer fabrication precision\cite{Portalupi_2011, Taguchi_2011, Thomas_2011} has brought about ultra-high-$Q$ cavity designs\cite{Akahane_2003a, Song_2005, Kuramochi_2006} with mode volumes close to the diffraction limit, as well as low-loss, slow-light engineered waveguides\cite{Baba_2008}. This, together with the recent experimental success of Purcell-enhancing the emission of a single dot in a PHC waveguide\cite{Viasnoff-Schwoob_2005, Lund-Hansen_2008, Sapienza_2010, Schwagmann_2011, Hoang_2012, Laucht_2012}, and even reaching the strong coupling regime in such a structure\cite{Gao_2013}, suggests that a PHC-QD system could be an ideal candidate for demonstrating photon-mediated excitation transfer between distant dots. 

Coherent interaction between two QDs at subwavelength distance in a microcavity has been recently observed\cite{Albert_2013}. At longer distance, the interaction was theoretically shown to be finite but weak (as compared to typical radiative loss and decoherence rates) in $3D$ (bulk)\cite{Parascandolo_2005} and $2D$\cite{Tarel_2008} spatially homogeneous dielectric environment. The ideal compromise between interaction strength and range is thus expected in a $1D$ environment like a PHC waveguide, and indeed, the possibility for entangled states between distant QDs coupled to such a structure has been demonstrated\cite{Yao_2009}, and the characteristic interaction distance was estimated\cite{Minkov_2013} to be given by $r_{12} = 2v_g/\gamma$, where $v_g$ is the group velocity at the exciton resonant frequency, while $\gamma$ is the loss rate of the waveguide modes. However, it is known that disorder residual in the fabrication process dramatically affects the slow-light guided modes. In Ref. [\onlinecite{Minkov_2013}], we partially took this into account by introducing a phenomenological loss rate $\gamma$ as stemming from disorder-induced (extrinsic) losses, while the assumption of a perfectly ordered PHC structure implied that the effect of Anderson localization of light\cite{Savona_2011} was not included. In this work, we simulate realistic systems with different magnitudes of the disorder, and show that while light localization indeed has a profound effect on both range and magnitude of the dot-dot excitation transfer rate, this latter is still sizable, compared to typical decoherence rates, even at several $\mu m$ distance.

\begin{figure*}[ht]
\centerline{\includegraphics[width=14cm, trim = 0.5in 0in 0in 0.8in, clip = true]{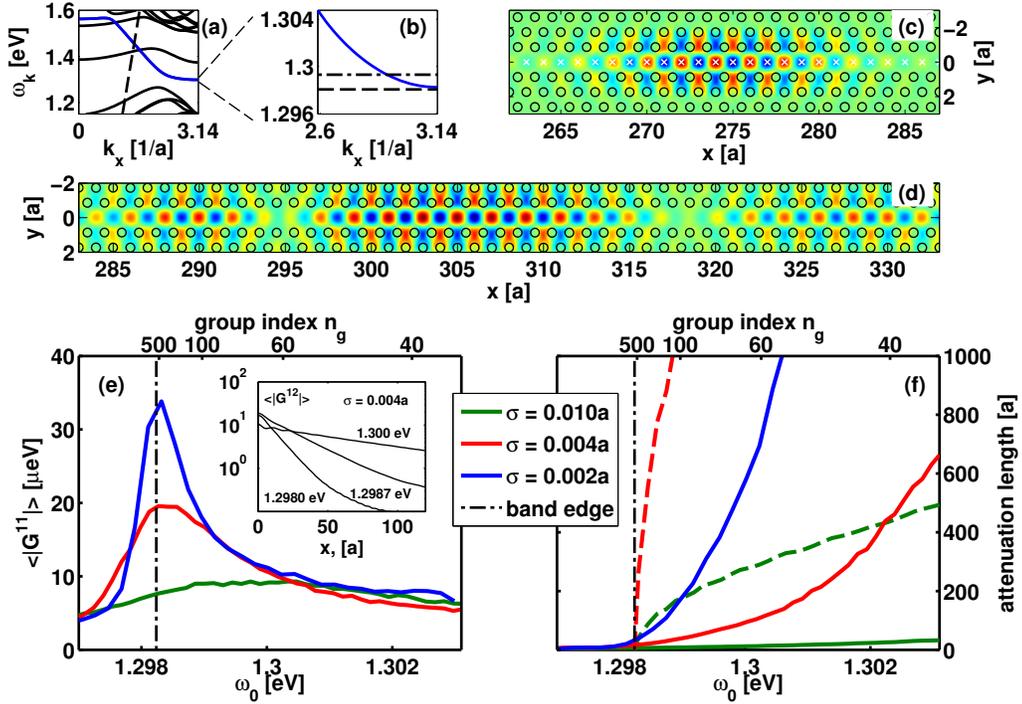}}
\caption{(Color online) (a): Band structure of the regular waveguide; the main guided band is shown with a blue line, while the light cone - with a dashed black line. (b): Zoom-in close to the guided band edge. (c) $y$-component of the electric field ($E_y$) of the mode at frequency shown by a dashed line in (b); the white crosses indicate the elementary cell centers, at which a quantum dot is potentially placed. (d) $E_y$ of the mode at frequency shown by a dashed-dotted line in (b). (e), (f): For three magnitudes of the fabrication disorder, (e): the average self-interaction (zero-distance) coupling term vs. $\omega_0$, and (f): the attenuation length of the distance-dependence, computed with (solid lines) and without (dashed lines, $=2v_g/\gamma$) localization effects. For $\sigma = 0.002a$, the $2v_g/\gamma$ curve cannot be distinguished from the band edge line on the scale of the plot. The inset in (e) shows the averaged $|G^{12}|$ vs. distance for three $\omega_0$-s (marked), and $\sigma = 0.004a$. }
\label{figure1}
\end{figure*}

The waveguide studied here is formed by a missing row of holes in a triangular lattice of circular holes etched in a dielectric slab suspended in air ($W1$ waveguide). The specific parameters, relevant to InGaAs quantum dots in GaAs structures\cite{Faraon_2008, Reinhard_2012}, are: lattice constant $a = 260 \mathrm{nm}$, hole radius $65 \mathrm{nm}$, slab thickness $120 \mathrm{nm}$, and a real part of the refractive index $\sqrt{\varepsilon_{\infty}}=3.41$. In the absence of fabrication disorder, the structure presents $1D$-periodicity along the direction of the missing holes, thus the modes are folded into Bloch bands (Fig. \ref{figure1} (a)). Everywhere below, we focus on the main guided band (blue line), in the spectral range close to the band edge (panel (b)). Fabrication disorder is introduced in the form of random fluctuations in the $x$ and $y$ positions and the radius of each hole, drawn from a Gaussian random distribution with zero mean standard deviation $\sigma$. A waveguide of length $512a$ is simulated, and in presence of disorder, its electromagnetic modes are computed by expansion on the basis of the Bloch modes of the regular structure\cite{Savona_2011, Savona_2012}.

Without disorder, guided modes in the considered spectral range are lossless, as they lie below the light-cone (dashed line of Fig. \ref{figure1}, panel (a)) and thus do not radiate outside the slab. Disorder has several important effects. First, it mixes those modes with the ones above the light-cone, introducing ``extrinsic'' losses, i.e. imposing a finite probability for out-of-plane radiation. Second, it limits the maximum group index, which in the ideal case goes to infinity at the band edge, and introduces modes that lie \textit{below} the band edge of the regular structure, i.e. the density of states of the disordered guide presents a Lifshitz tail below the van Hove singularity\cite{Savona_2011, Huisman_2012}. In addition, disorder induces Anderson localization of light\cite{Topolancik_2007, Sapienza_2010, Savona_2011}, which for states close to or below the band edge can be extremely strong (Fig. \ref{figure1} (c)), localizing the electric field over several elementary cells. The field profiles of such modes resemble those of PHC cavities, and both strong Purcell enhancement\cite{Sapienza_2010} and cavity-like vacuum Rabi splitting\cite{Gao_2013} of a single QD coupled to such a mode has already been observed.  Modes slightly higher in frequency become more extended, and present more than one lobes (panel (d)), and in fact provide the ideal compromise between strength and range of the dot-dot excitation transfer. In this work, we always consider two dots in the waveguide, which are placed in the center of an elementary cell (at a position indicated by a white cross in Fig. \ref{figure1} (c)), and so at a distance multiple of $a$ from each other.

To quantify the QD-W1 and the effective QD-QD coupling, we use the Green's function formalism that we developed in Ref. [\onlinecite{Minkov_2013}] starting from Maxwell's equations for the PHC with an added linear susceptibility due to the QDs (valid in the low-excitation regime). The effective coupling strength is 

\begin{equation}
G^{12}(\omega_0) =  d^2 \frac{2 \pi}{\epsilon_{\infty} \hbar} \frac{\omega_0^2}{c^2} G(\mathbf{r}_{\alpha}, \mathbf{r}_{\beta}, \omega_0), 
\label{G12}
\end{equation}

where $d$ is the dipole moment of the dot, $\epsilon_{\infty}$ is the dielectric constant of the semiconductor, $\omega_0$ is the exciton resonance frequency, and $G(\mathbf{r}_{1}, \mathbf{r}_{2}, \omega_0)$ is the photonic Green's function at the dot positions, computed here using the resolvent representation once the orthonormal set of electric field modes of the waveguide is obtained through the Bloch-mode expansion. An important remark is thus that Eq. \ref{G12} takes into account the \textit{many-mode} exciton-photon coupling that is bound to occur close to the band edge, where the density of photonic modes is high. The dipole moment $d$ can be estimated from the spontaneous emission half-life of the dot embedded in bulk semiconductor, which was taken here to be $\Gamma = 1 \mathrm{ns}$. In the weak-coupling regime, the Purcell enhancement factor for a single dot in the PHC is related to the zero-distance coupling as $PF = -2\Im(G^{11}(\omega_0))/\Gamma$. More generally, $G^{11}(\omega_0) = \sum_{m} |g_m|^2/(\omega_m-i\gamma_m - \omega_0)$, where the sum runs over all the electromagnetic field eigenmodes, $g_m$ is the coupling rate of the dot to each mode, while $\omega_m$ and $\gamma_m$ are respectively the frequency and loss rate of each mode. If the coupling rate exceeds the loss rates, strong coupling sets on and, in the case of both one and two dots, the irreversible radiative decay is replaced by an  oscillatory dynamics of the energy transfer\cite{Minkov_2013}. In this sense, $G^{12}(\omega_0)$ is a measure of the frequency of this oscillatory excitation transfer process between two distant dots. 

It should be noted that, while localized modes always appear in the presence of disorder, their particular shape, and the position of the localized lobes, differs vastly among disorder realizations. Thus, here we perform the analysis using a \textit{configuration average} over 400 different realizations of the waveguide disorder, and a \textit{running average} over the position of the first dot in each particular waveguide. The dependence with inter-dot distance of the averaged magnitude of the excitation transfer rate $\langle |G^{12}| \rangle$ is shown in the inset of panel (e) of Fig. \ref{figure1} for three different exciton transition frequencies ($\omega_0 = 1.2980 eV$, $\omega_0 = 1.2987 eV$ and $\omega_0 = 1.3000 eV$, with band edge at $\omega = 1.2982 eV$) and for $\sigma = 0.004a$. These plots show some deviation from an exponential law at large distances, but this is an unphysical result originating form the finite size of our simulation domain, and occurs at very small values of $G^{12}$ which are scarcely relevant to our conclusions. For each $\omega_0$, an exponential function can thus be fitted in the region where the decay is a straight line on a logarithmic plot, and an attenuation length can be extracted. On this basis, panels (e) and (f) give detailed information about the dot-dot interaction for three different disorder magnitudes. The strength is quantified in panel (e), through the averaged zero-distance term  $\langle |G^{12}| \rangle$, while the range -- in panel (f) through the interpolated attenuation length. Finally, even though the group index $n_g$ cannot be well-defined in the presence of localization, its value in the ideal-PHC case is given on the top $x$-axis in both panels.

Some previous experimental works\cite{Sapienza_2010, Gao_2013} in which single-dot coupling to a PHC waveguide has been demonstrated take advantage of large PHC disorder as a means to have strongly localized modes. That this is beneficial is not directly obvious from the large-disorder result shown in Fig. \ref{figure1} (e), which never exceeds $10 \mathrm{\mu eV}$. It should be kept in mind however that this result represents the configuration-averaged zero-distance coupling. In few individual configurations in which the dot is sitting exactly on top of a strongly localized mode, the same coupling can exceed $100 \mathrm{\mu eV}$. In any case, such a strong disorder makes it very unlikely to have long-distance dot-dot interaction, as can be seen from panel (f). For $\sigma = 0.004a$ ($\approx 1 nm$ in typical systems, realistically achievable), however, the attenuation length becomes sizable -- in the order of $100a$ which corresponds to the $10~\mathrm{\mu m}$ range. Notice, though, that the localization still has a drastic effect as compared to the case of extrinsic losses only, where the transfer rate is determined by the ratio $2 v_g/\gamma$ plotted as dashed lines in panel (f), that was analyzed in Ref. [\onlinecite{Minkov_2013}]. In the figure, $\gamma$ was taken as the average over the loss rates, computed through Bloch-mode expansion, for each of the disorder magnitudes, and corresponds to a quality factor of $\mathcal{Q} = 95 000$ for each mode in the $\sigma = 0.004a$ case. The drastic influence of Anderson localization emerges in the fact that modes at a given frequency are characterized by a localization length which also determines the spatial decay of the light transport process at that frequency. The corresponding decay length is generally much smaller than that associated to \textit{ballistic} propagation in presence of a phenomenological extrinsic loss rate, as studied in Ref. [\onlinecite{Minkov_2013}]. In the context of light propagation in PHCs, the effect of Anderson localization is often referred to as \textit{backscattering losses},\cite{Hughes_2005, OFaolain_2007, Mazoyer_2010} and was shown to severely degrade the transmission for frequencies close to the band edge. 
 
\begin{figure}[ht]
\centerline{\includegraphics[width=8cm, trim = 0.3in 0in 0in 0.8in, clip = true]{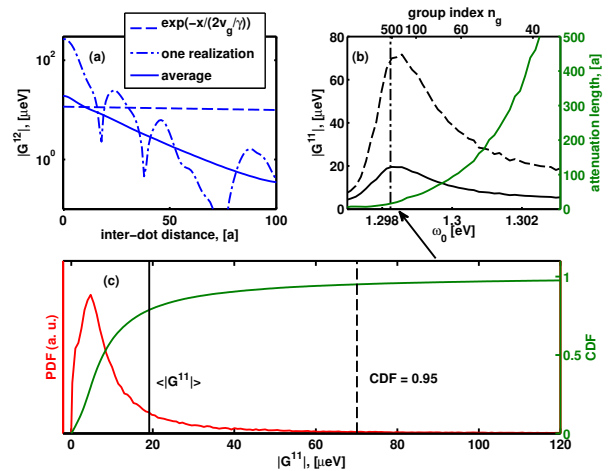}}
\caption{(Color online) (a) Excitation transfer rate vs. distance without light localization (dashed), or with, for a single disorder realization (dashed-dotted) and the configuration average (solid), for $\sigma = 0.004a$ and $\omega_0 = 1.2985 \mathrm{eV}$ (indicated by an arrow in (b)). (b) The two solid lines are the same as the $\sigma = 0.004a$ lines in panels (e) and (f) of Fig. \ref{figure1}; the dashed line shows the value of $|G^{11}|$ for which the CDF (panel (c)) is 0.95. (c): PDF and CDF of $|G^{11}|$, with $\langle |G^{11}| \rangle$ and the CDF = 0.95 values explicitly indicated.}
\label{figure2}
\end{figure}

While the configuration average gives a good estimate of the interaction strength, it is also important to understand the underlying statistics, to know what one could expect in an actual experiment. In Fig. \ref{figure2} (a), we compare the configuration averaged interaction to a single realization and to the $\exp(-x/(2v_g/\gamma))$ dependence, for $\omega_0 = 1.2985 \mathrm{eV}$ (indicated by an arrow in the figure). On the scale of the plot, the latter appears as a horizontal line, illustrating again the difference that localization effects make. When those are taken into account (solid line), the interaction range is suppressed but the magnitude at short distances is increased -- which is natural given the presence of cavity-like modes. The single-realization line was taken for one particular disorder configuration and for a position of the first dot at which the zero-distance coupling is strong - exceeding $100 \mathrm{\mu eV}$, illustrating that the statistics present a large variance. Indeed, the probability density function (PDF) of $|G^{11}|$ exhibits a very long tail towards high values (Fig. \ref{figure2} (c)). Thus, in panel (b), we compare the averaged value of $|G^{11}|$ that was given in Fig. \ref{figure1}, to the value for which the cumulative distribution function (CDF) is equal to $0.95$. Put simply, the dashed line in Fig. \ref{figure2} (b) gives the interaction magnitude that one can expect from one in every 20 samples. It is then clear that  $|G^{11}|$ can exceed $70 \mathrm{\mu eV}$, and a value of above $30 \mathrm{\mu eV}$ can be expected even for frequencies for which the interaction range is of the order of $100a$. 
 
\begin{figure}[t]
\centerline{\includegraphics[width=8cm, trim = 0.5in 0in 0in 0in, clip = true]{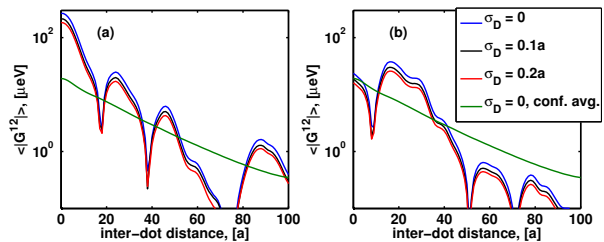}}
\caption{(Color online) (a), (b): For two different positions of the first dot, the excitation transfer rate vs. distance with the dots placed exactly in the center of an elementary cell (blue line) and with some positioning disorder (black and red lines, averaged over dot positioning). The configuration average over PHC disorder is also shown (green line).}
\label{figure3}
\end{figure}

It is important to note that the PHC disorder plays a more important -- and non-trivial -- role than the imperfect positioning of the quantum dots. This is illustrated in the two panels of Fig. \ref{figure3}: the green lines show the configuration-averaged interaction for $\sigma = 0.004a$ at the same frequency used in Fig. \ref{figure2}, $\omega_0 = 1.2985 \mathrm{eV}$, while the blue curves are two different specific realizations. When disorder in the positioning of the QDs is introduced, with standard deviation $\sigma_D$, the interaction is simply scaled down by a constant not far from unity. The observed tendency is expected, since the electric field does not vary strongly on the length-scale of a few tens of nanometers.

In conclusion, in this work we investigated the possibility for excitation transfer between distant quantum dots in a photonic crystal waveguide. Due to Anderson localization of light, disorder in the position and size of the PHC holes was found to have a highly non-trivial effect on the interaction, thus statistics based on $400$ different disorder realizations were analyzed. For $\sigma = 0.004a$, the averaged excitation transfer rate was found to be larger than $10 \mathrm{\mu eV}$ at distances in the order of $10 \mathrm{\mu m}$. In addition, in a 1-out-of-20 setting, the rate reaches $50 \mathrm{\mu eV}$ and more. The transfer time to which this corresponds is of the order of $10 \mathrm{ps}$ -- close to the single-qubit operation time and much shorter than the decoherence time measured in these systems. Disorder in the positioning of the dots has a straightforward scaling down effect, which is small when the precision is in the range of tens of nanometers. This shows that a PHC waveguide – or a similar structure engineered to further enhance the dot-dot coupling – is an ideal candidate for establishing a long-distance, photon-mediated QD state transfer.

\begin{acknowledgments}
This work was supported by the Swiss National Science Foundation through Project No. $200020\_132407$.
\end{acknowledgments}

\bibliographystyle{apsrev}
\bibliography{polaritons_paper}

\end{document}